\documentclass[journal]{vgtc}                

\ifpdf
  \pdfoutput=1\relax                   
  \usepackage{graphicx}                
  \DeclareGraphicsExtensions{.pdf,.png,.jpg,.jpeg} 
\else
  \usepackage{graphicx}                
  \DeclareGraphicsExtensions{.eps}     
\fi

\graphicspath{{figures/}{pictures/}{images/}{./}} 

\usepackage[dvipsnames,table, svgnames]{xcolor}
\usepackage{tabu}                      
\usepackage{booktabs}                  
\usepackage[normalem]{ulem}
\usepackage{textgreek} 
\usepackage{eurosym}
\usepackage{ccicons}
\usepackage{xspace}
\usepackage{cuted}
\usepackage{flushend}
\usepackage{graphicx} 
\usepackage{multirow}

\useunder{\uline}{\ul}{}
\newcommand{\eg}{e.\,g.}
\newcommand{\ie}{i.\,e.}

\definecolor{antiquefuchsia}{rgb}{0.57, 0.36, 0.51}

\newlength{\picturewidth}
\newlength{\pictureheight}

\onlineid{1229}

\title{Data Embroidery with Black-and-White Textures}

\author{%
  \authororcid{Tingying He}{0000-0002-9670-5587},
  \authororcid{Petra Isenberg}{0000-0002-2948-6417},
  \authororcid{Tobias Isenberg}{0000-0001-7953-8644}
}

\authorfooter{
\item Tingying He (\raisebox{-.5pt}{\includegraphics[height=6pt]{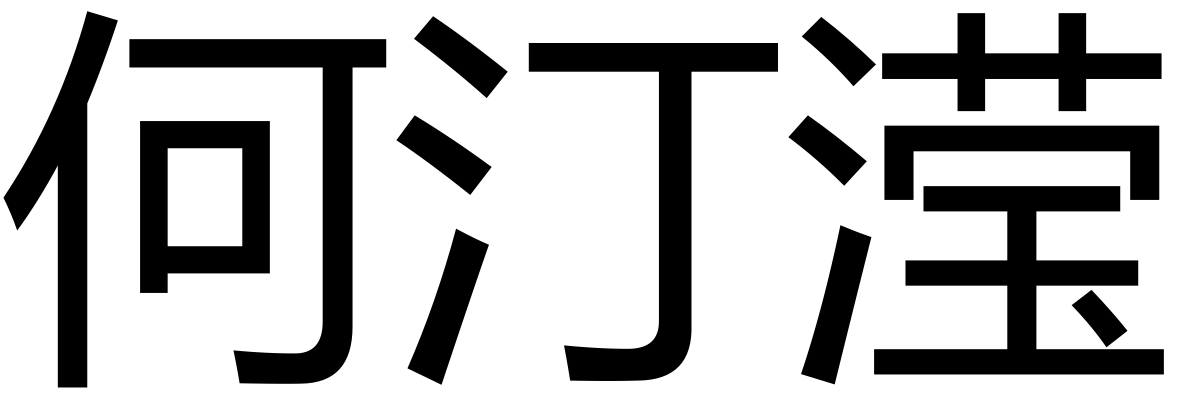}}), Petra Isenberg, and Tobias Isenberg are with Université Paris-Saclay, CNRS, Inria, LISN, France. E-mail: \{tingying.he\,$|$\,petra.isenberg\,$|$\,tobias.isenberg\}@inria.fr.}

\shortauthortitle{He \MakeLowercase{\textit{et al.}}: Data embroidery with black-and-white textures}

\abstract{%
We investigated data embroidery with black-and-white textures, identifying challenges in the use of textures for machine embroidery based on our own experience. Data embroidery, as a method of physically representing data, offers a unique way to integrate personal data into one's everyday fabric-based objects. Owing to their monochromatic characteristics, black-and-white textures promise to be easy to employ in machine embroidery. We experimented with different textured visualizations designed by experts and, in this paper, we detail our workflow and evaluate the performance and suitability of different textures. We then conducted a survey on vegetable preferences within a family and created a canvas bag as a case study, featuring the embroidered family data to show how embroidered data can be used in practice.
} 


\CCScatlist{ 
 \CCScat{K.6.1}{Management of Computing and Information Systems}%
{Project and People Management}{Life Cycle};
 \CCScat{K.7.m}{The Computing Profession}{Miscellaneous}{Ethics}
}

\teaser{
\setlength{\pictureheight}{32mm}
    \centering
    \vspace{1em}
    \footnotesize%
    (a)~\includegraphics[height=0.27\linewidth]{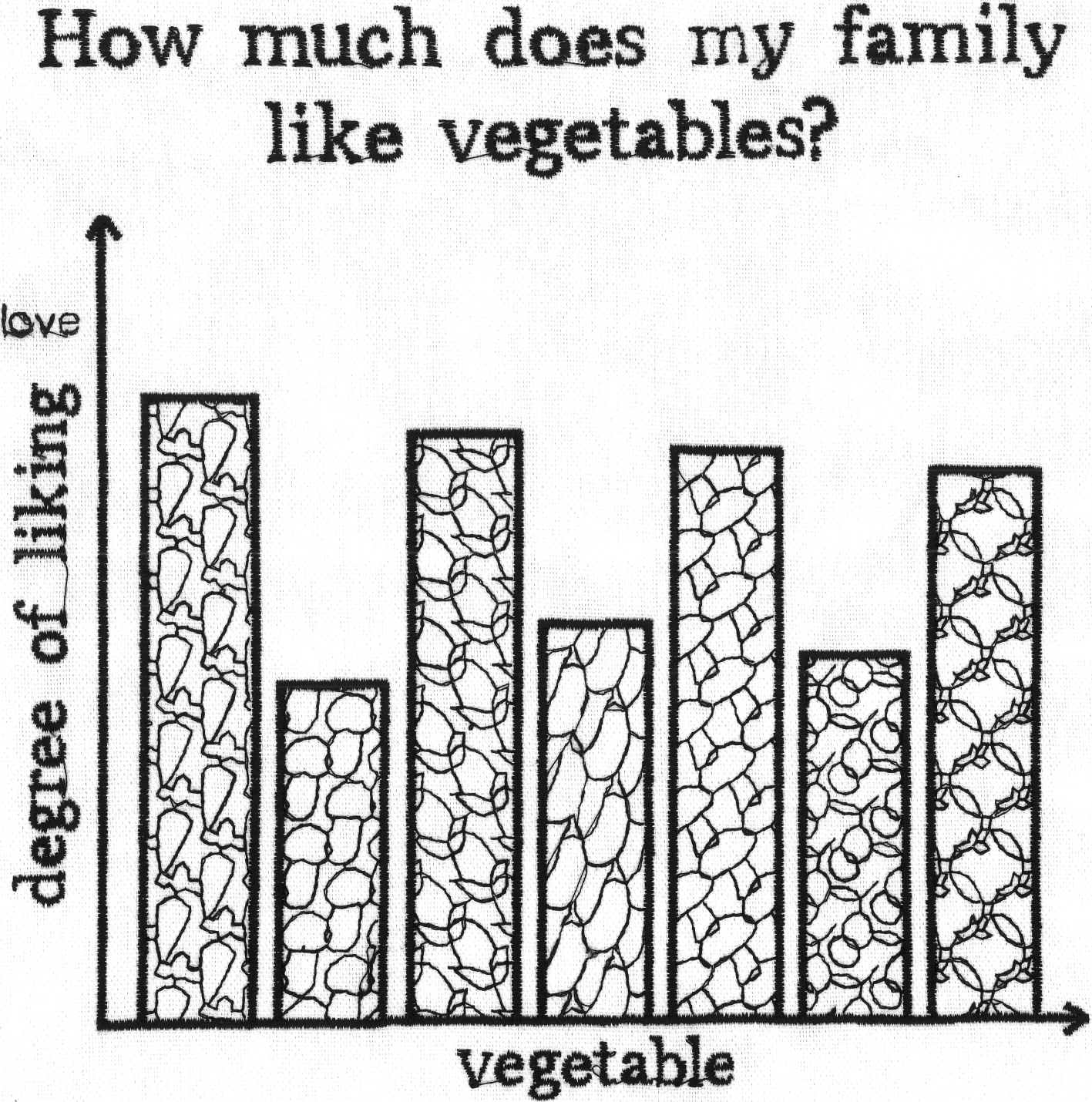}\hfill
    (b)~\includegraphics[height=0.27\linewidth]{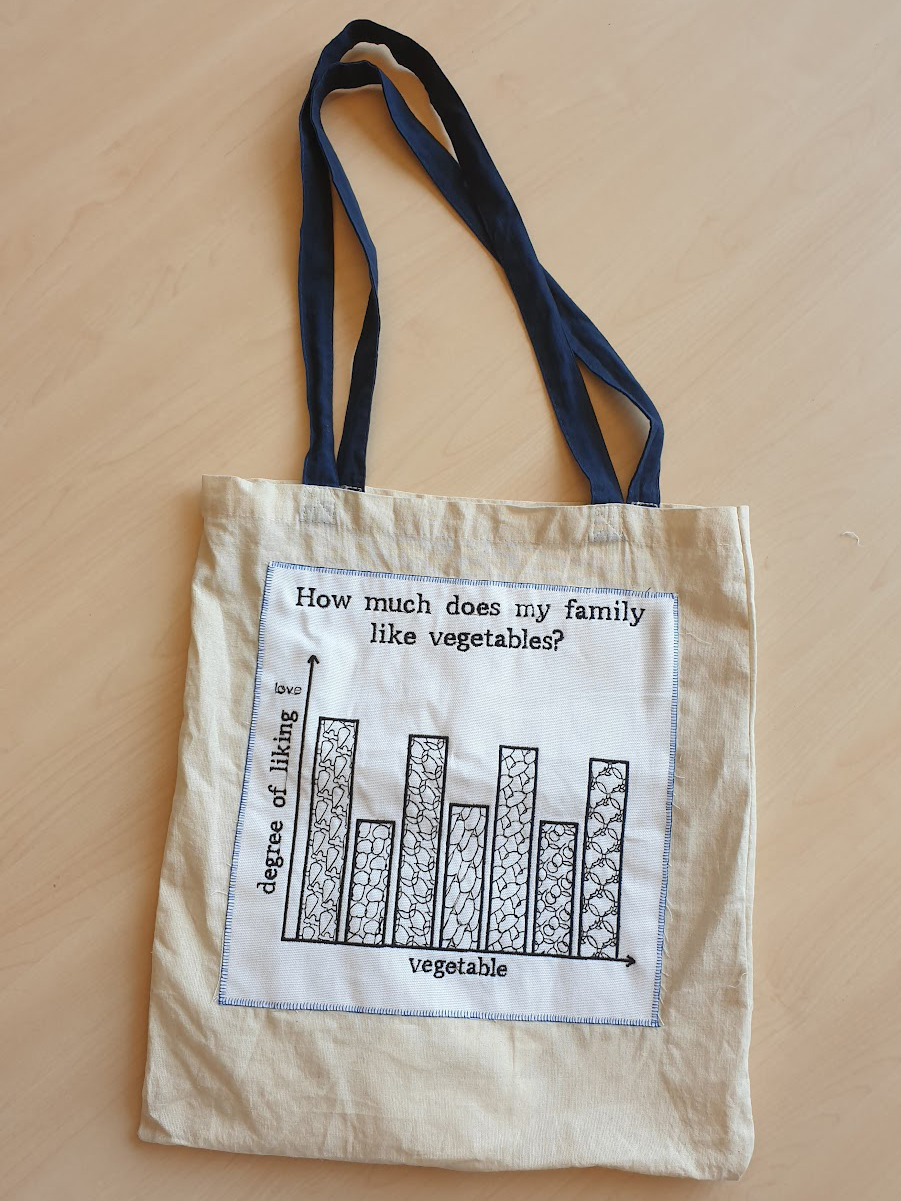} \hfill
    (c)~\includegraphics[height=0.27\linewidth]{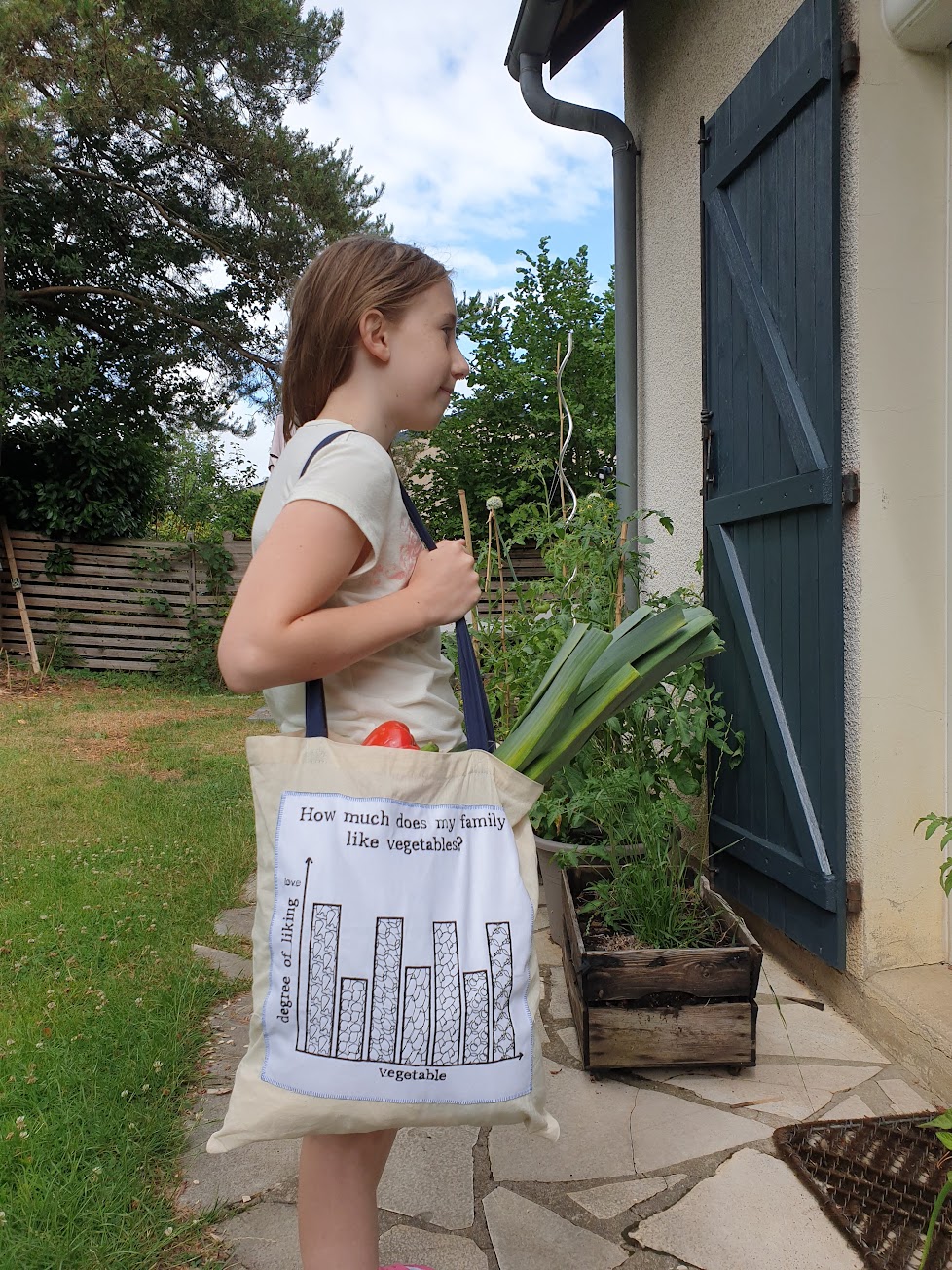}\hfill
    (d)~\includegraphics[height=0.27\linewidth]{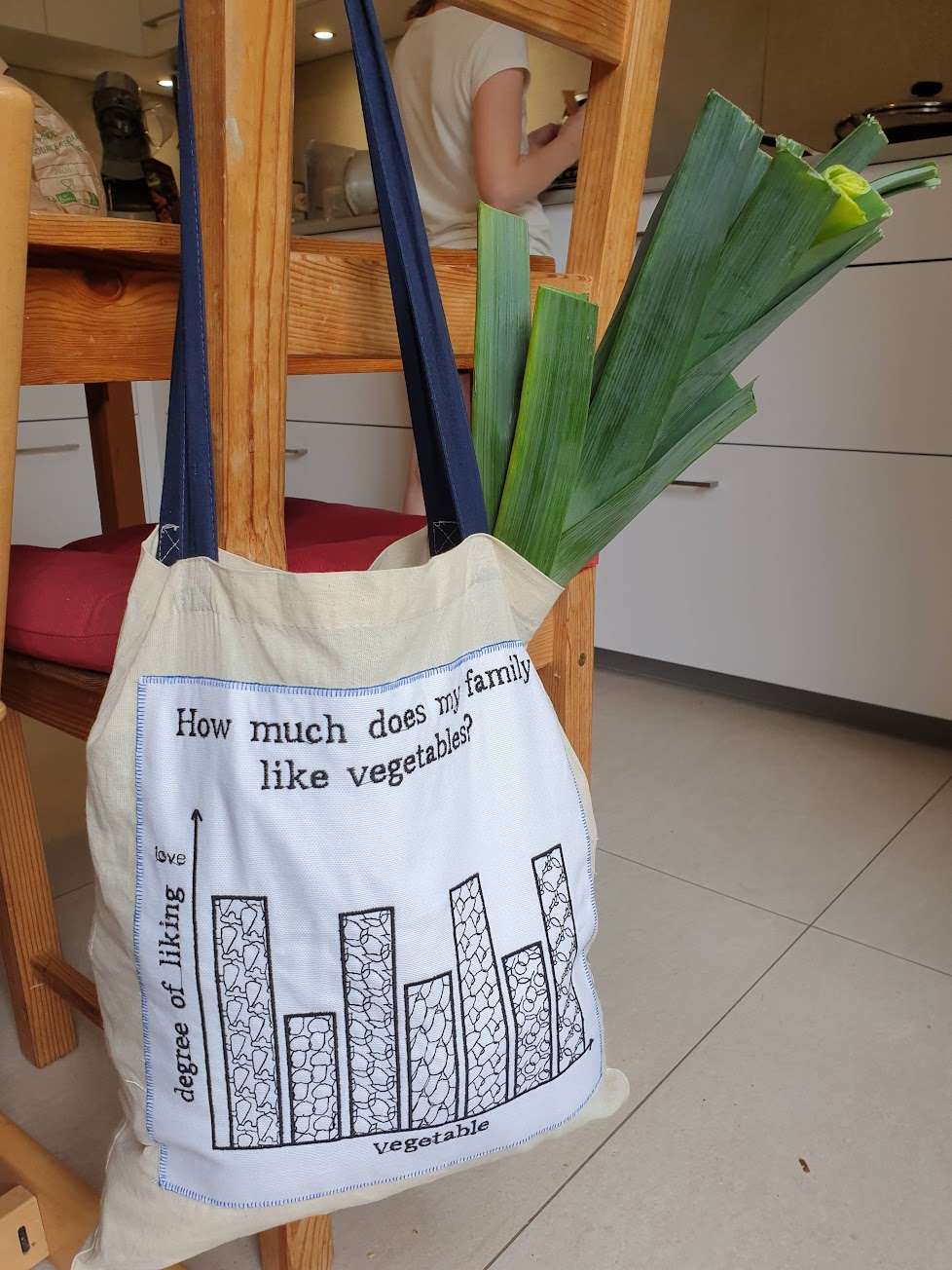}
    \caption{(a): An embroidered chart with black-and-white textures displaying the results of a survey within a family. (b): A canvas bag featuring the embroidered chart on the left. (c) and (d): the bag being used within the family. }
    \label{fig:teaser}
}

\newcommand{\inlinevis}[3]{\raisebox{#1}[0pt][0pt]{\includegraphics[height=#2]{#3}}}

\newlength{\plotheight}
\manuscriptnote{\vspace{-2.8em}}

\begin{document}



\noindent\begin{minipage}{\textwidth}
\centering
\includegraphics[width=0.9\linewidth]{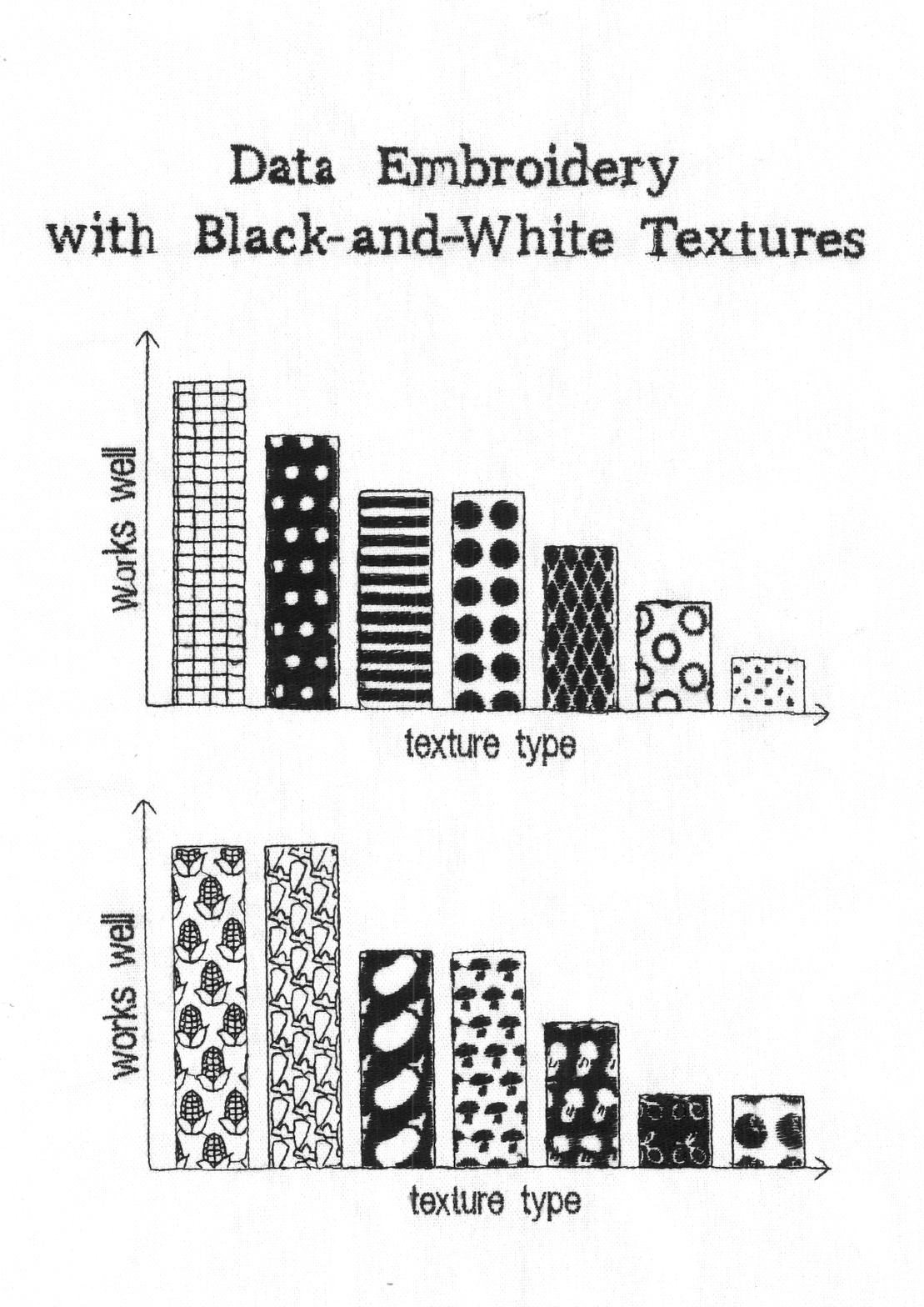}
\end{minipage}
\newpage

\maketitle


\section{Introduction}
\label{sec:introduction}

Data embroidery \cite{wannamaker:2019:data} is an innovative technique for data physicalization \cite{Jansen:2015:opportunities}. Machine embroidery as a computer-numerically controlled (CNC) technology makes it  possible to produce complex data embroideries (relatively) quickly and integrate them into fabric-based personal belongings \cite{wannamaker:2019:data}. Data embroidery of personal data has potential because a less conventional approach to visualization may stimulate people to explore their own data more intensively \cite{wang:2015:design}. It can also serve as an ambient visualization within a home setting, thereby initiating dialogues with curious visitors \cite{Pousman:2007:casual}. Data embroidery can, like in our case, be accessible to a broad set of the population through local Fablabs. 

A promising yet so far unexplored avenue within data embroidery involves the use of black-and-white textures. Before the ubiquity of color printing, these monochromatic textures served as a powerful visual channel for data visualization (\eg, see the OldVisOnline collection \cite{Zhang:2024:OCD}). Their inherent simplicity facilitates the conversion of images to embroidery files, overcoming challenges associated with the lower color resolution of embroidery machines compared to color screens. Moreover, they eliminate the need for multiple color changes during the embroidery process, enhancing efficiency.

In our work we explore data embroidery with black-and-white textures and contribute the following: (1) a detailed and hands-on workflow for creating data embroidery from an existing black-and-white textured chart image, (2) preliminary evaluations of which textures can be most effectively translated into data embroidery, and (3) a showcase of data embroidery of personal data with black-and-white textures---a canvas bag with an embroidered chart visualizing a within-family survey.
\section{related work}
\label{sec:related_work}
Many researchers and artists have experimented with embroidery as a means to create visualization artwork. For instance, Liz Bravo \cite{LizBravo} manually embroidered charts of the distribution of U.S. cotton from 1942 to 1948, a visualization originally created by Mary Eleanor Spear, a pioneer in data visualization. Olivia Johnson \cite{OliviaJohnson} used cross-stitch techniques to create charts about gender inequality and discrimination at workplaces \cite{cabric:2023:eleven}. Hand embroidery has also been used to explore personal data. For example, Jane Zhang \cite{JaneZhang} logged her anxiety over 21 days and created an embroidered chart to visualize it---akin to other efforts in personal data visualization \cite{Posavec:2016:DD}. In the visualization research literature, Smit \cite{Smit:2021:DataKnitualization} has expanded the fabric-based data visualization landscape by exploring hand knitting as a potential medium for data physicalization, resulting in several \emph{data knitualization} works. All these efforts, however, have employed manual methods to represent data on fabric, while we explore a more automated process.

Machine embroidery has gradually started to draw researchers' attention as well. Wannamaker et al. \cite{wannamaker:2019:data} explored the use of CNC embroidery machines for expressing personal data and embroidered a personal data physicalization representing text message data on a blanket. Schneider \cite{DataVisualizationWithMachineEmbroidery} provided a tutorial on data visualization with machine embroidery using Ink/Stitch \cite{inkstitch}, in which he outlined a general workflow of computerized embroidery. This workflow notably includes an essential step of reducing colors of the drawing to adapt it to the embroidery constraints, which underscores the suitability of monochromatic charts for machine embroidery. We used this process as an inspiration but specifically focused on textured visualizations.

\section{Machine Embroidery for Visualization}
\label{sec:embroidery}
We first detail the workflow we followed for creating data embroidery from a chart image, as well as the choices we made during this process. We then discuss the common issues we encountered during our data embroidery process and possible troubleshooting methods.

\subsection{Machine embroidery process}
\label{sec:machine-embroidery-process}
\begin{figure}
    \centering
		\includegraphics[width=1\columnwidth]{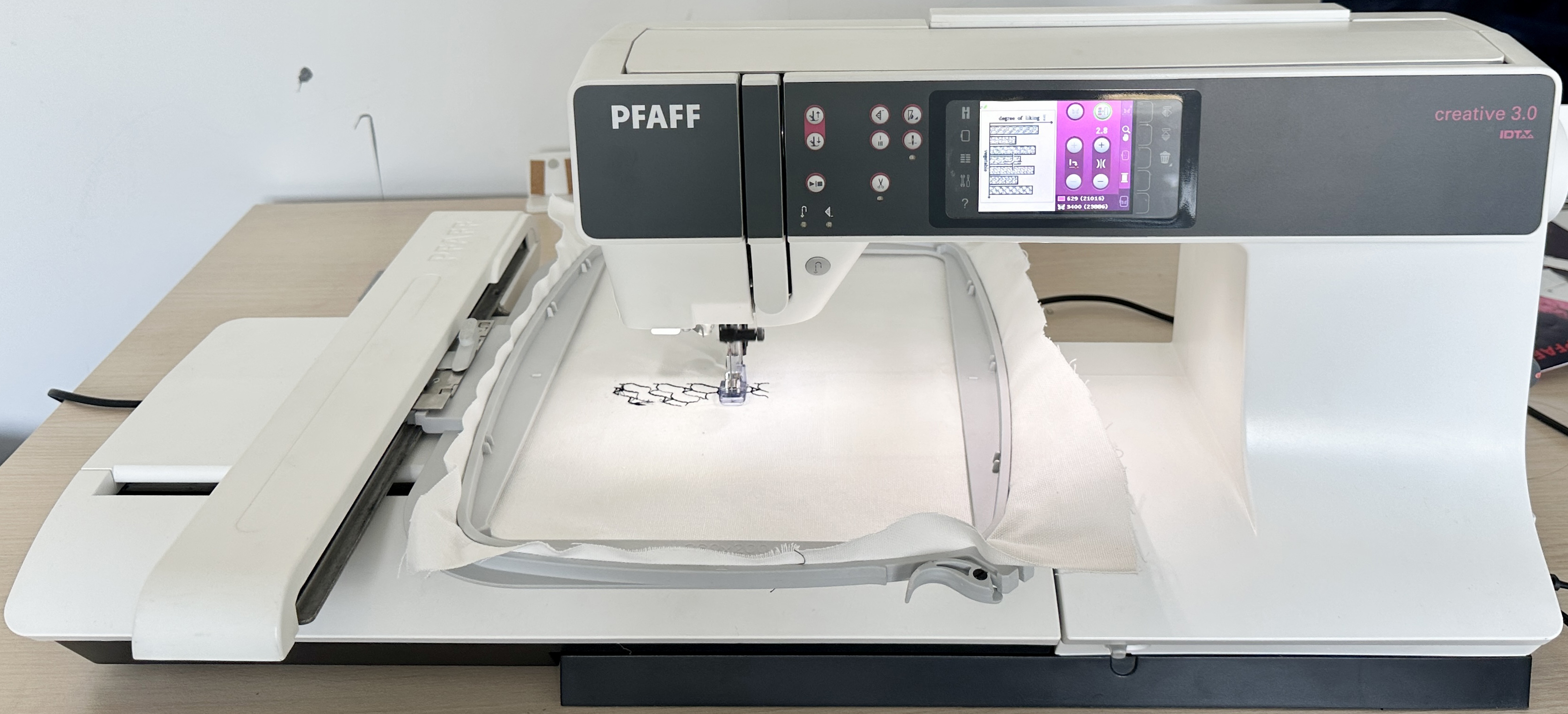}	
    \caption{The embroidery machine is embroidering a chart. Notice the hoop that holds and stretches the fabric for the embroidery process.}
    \label{fig:machine_embroidery}
\end{figure}

Embroidering a data representation, much like any image, consists of two significant steps: (1) the preparation of an embroidery file that describes the path of the needle and other settings and (2) the actual embroidery process. In our work, we used the Pfaff Creative 3.0 Sewing and Embroidery Machine (\autoref{fig:machine_embroidery}), which affects further choices below.

\subsubsection{Embroidery File Preparation}
For the first step we selected mySewnet, a set of software tools recommended by Pfaff. Notably, numerous other free and open-source software tools and libraries are available as well, such as Ink/Stitch \cite{inkstitch}. Regardless of the chosen software, the procedure is similar: we open an image and process it within the embroidery software to then save it in an embroidery format. The specific input and output formats vary, depending on the capabilities of the embroidery software and machine. In our case, we used the pixel-based \texttt{PNG} format as the source and \texttt{VP3} (a format used by Pfaff machines) as the export format.

MySewnet provides three methods for converting an image to an embroidery file: converting the entire image, tracing the elements’ outline, or tracing the image border. To achieve optimal results, we divided the chart image into three components---areas and two types of lines. We used the ``whole image convert'' method for areas, and the ``trace outline convert'' method for lines. The embroidery software, however, treats the entire image as a whole during the conversion process, meaning that certain parts can potentially impact others. For instance, if we would attempt to convert an entire bar chart with a line-based texture in one go, the line texture within the bar may distort the bar's outline. Consequently, we recommend the separation of the outline and the filling before the conversion process---even if both are based on lines. In the end, we thus processed area fillings, line-based texture fillings, and element outlines as separate parts and then reassembled them within the processing software, which lead to improved results.

After the conversion, we performed minor edits on the embroidery file in the tool mySewnet Digitilizing, for example we converted some curved stitches to straight lines based on the original input. In addition, we added text elements to the chart within the tool mySewnet Embroidery, because directly converting text from images often yielded bad results. We also adjusted the size of each component for a better fit with respect to each other. After finalizing the embroidery file, we imported it into the embroidery machine via USB, adjusted its positioning within the embroidery hoop (the frame that holds and stretches the fabric), and prepared for the physical embroidery process.

\subsubsection{Embroidery Process}
Before starting the embroidery process, we prepared the necessary materials. Through several small-scale experiments with various fabrics such as paper, cotton, linen, soft shell, and canvas, we found that canvas produced the best results for embroidering our textured charts. We thus decided to use canvas with a 2.5\,oz cut-away stabilizer (the material to put on the back of the fabric during the embroidery process), which is a recommended type of stabilizer for canvas. We ironed the fabric, bonded the fabric to the stabilizer with 505 temporary adhesive spray, and placed the fabric into the hoop. Next, we attached the hoop to the embroidery machine. We experimented with different needle and thread combinations, and finally used a 75/11 embroidery needle with polyester black thread for the best performance. After attaching the needle to the machine, we wound the bobbin (the part that holds the bottom thread) with the thread and threaded the needle. After these preparations we could initiate the embroidery process.

\subsection{Troubleshooting}
To get a good embroidery, it is important to ensure that the needle, the threading, and the bobbin are correctly set. A flawless setup, however, does not necessarily guarantee an error-free embroidery process. During our experimentation we faced several common issues as follows.

\begin{figure}
    \centering
		(a)~\includegraphics[width=0.45\columnwidth]{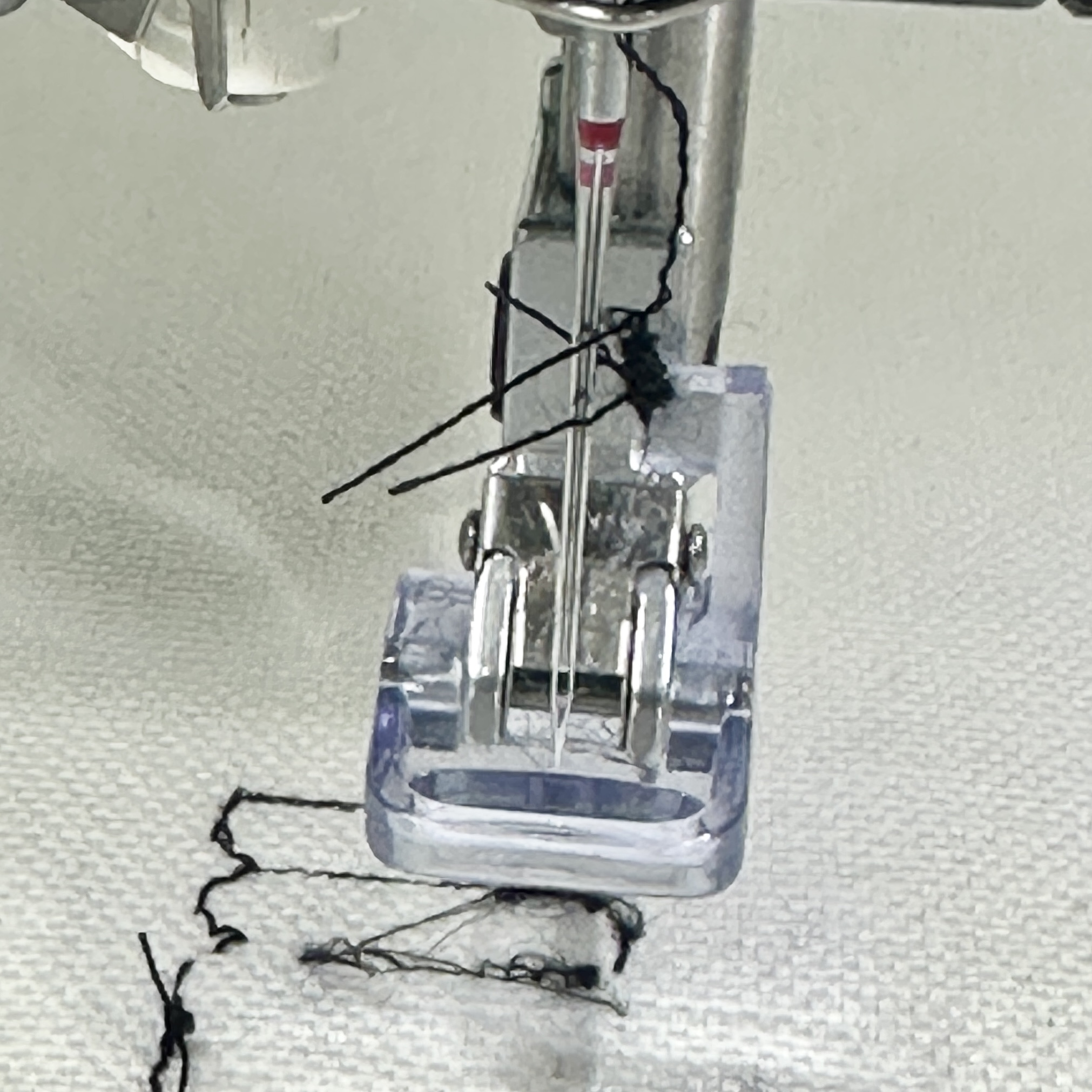}\hfill%
		(b)~\includegraphics[width=0.45\columnwidth]{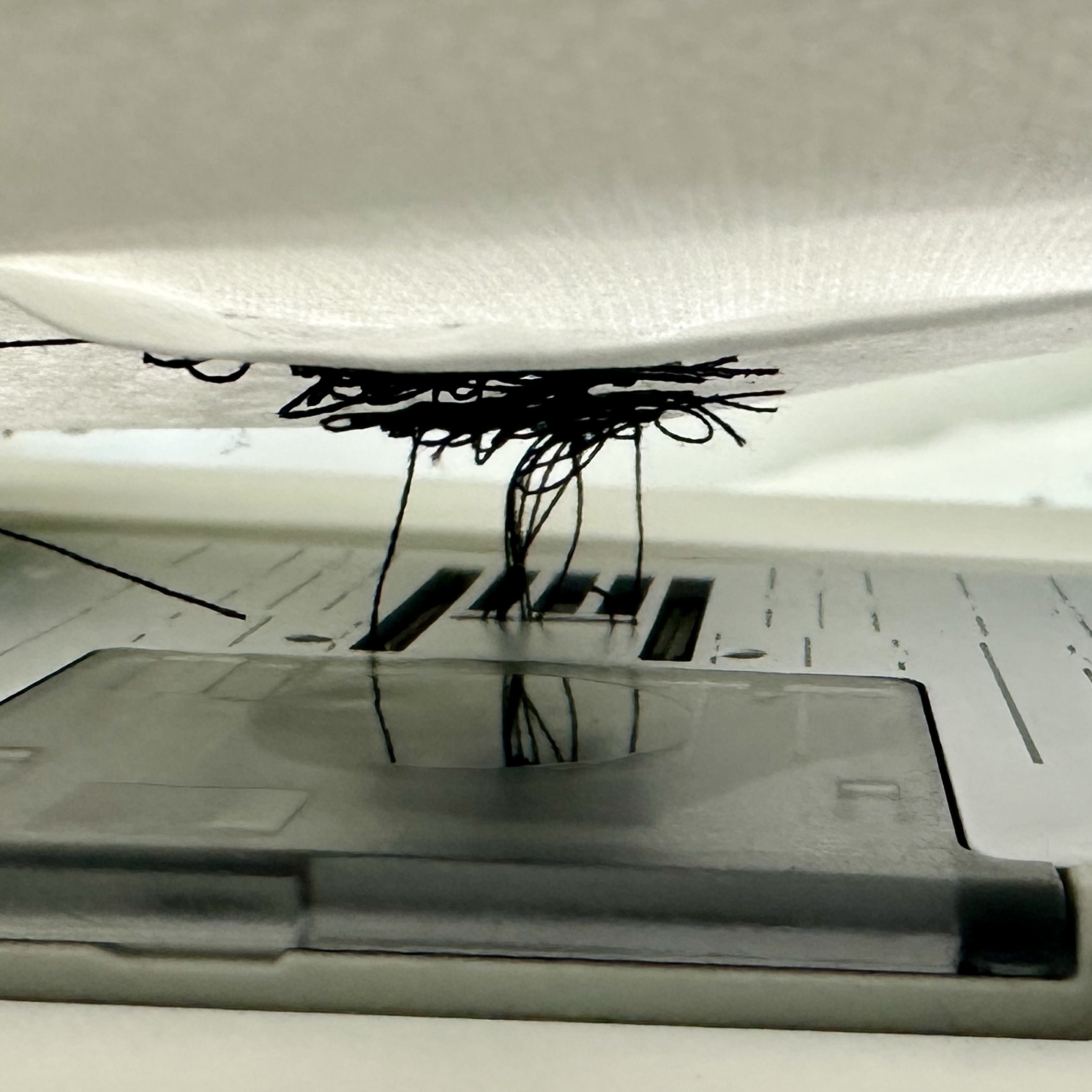}		
    \caption{Errors during the embroidery process: (a) thread breakage and (b) bobbin thread getting stuck under the needle plate.}
    \label{fig:embroidery_errors}
\end{figure}

A common issue we frequently encountered was \textbf{thread breaking during embroidery}, as shown in \autoref{fig:embroidery_errors}(a). If this happens it is essential to rethread the machine correctly. If thread breakage occurs frequently despite the right setup, however, one may need to consider reducing the embroidery speed. This reduction might decrease the stress on the thread and, subsequently, prevent breakage.

Another typical problem was the \textbf{bobbin thread getting stuck under the needle plate}, often accompanied by bobbin bunching (\autoref{fig:embroidery_errors}(b)). We then removed the needle plate and cleaned the tangled thread under the needle plate and around the bobbin. After clearing the jam, we rethreaded the bobbin to continue the process.

When noticing any error during embroidery, it is crucial to stop the process immediately, so observing the machine while it works is recommended. The machine may also automatically stop when it detects a problem. Before restarting, it is important to check for any missed stitches due to the error and to patch potentially missing stitches using the embroidery machine's ``step through stitch by stitch'' function.

\section{Which textures perform better for embroidery?}

The quality of data embroideried is influenced by several factors. Aside from the machine setup and the used materials, the pattern to be embroidered---in our case, the texture used in the design---can also impact the results. Despite having an optimal set-up, some textures are inherently more challenging to embroider than others, leading to more issues during the embroidery process. 

\begin{figure}
    \centering
		\includegraphics[width=0.47\columnwidth]{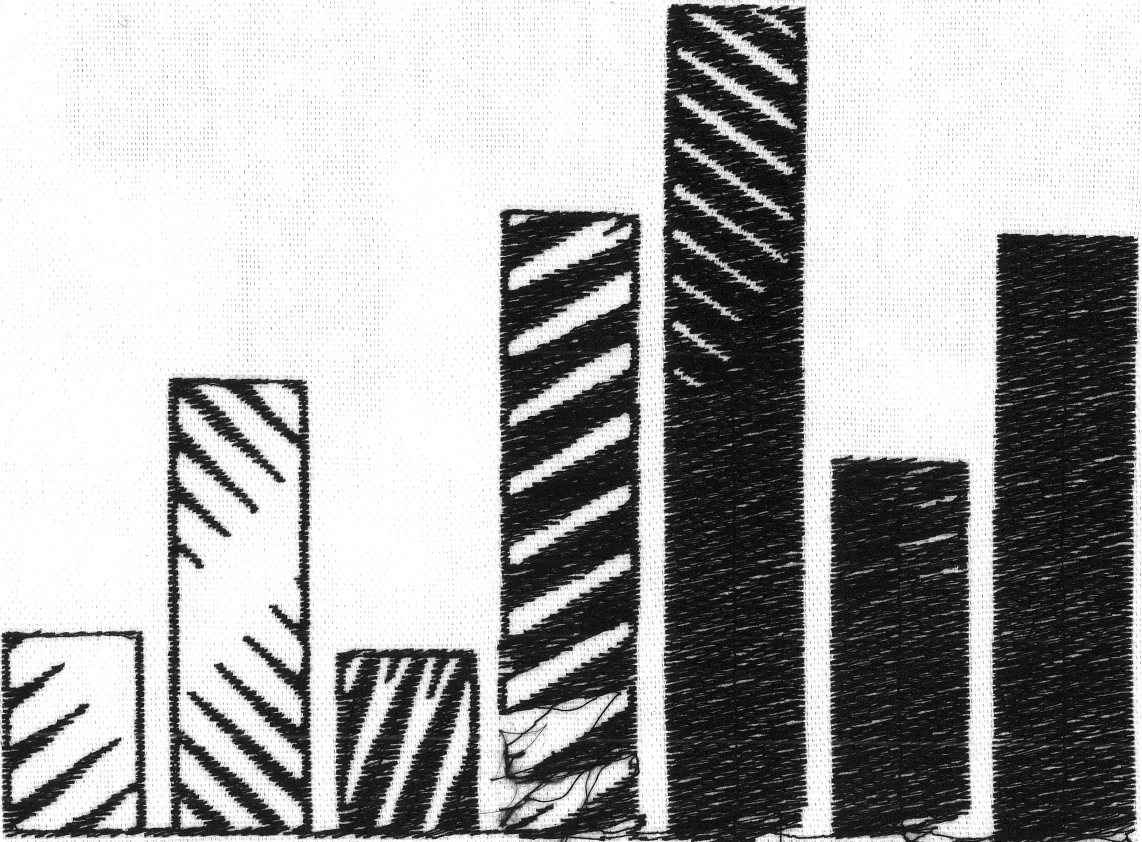}\hfill%
		\includegraphics[width=0.47\columnwidth]{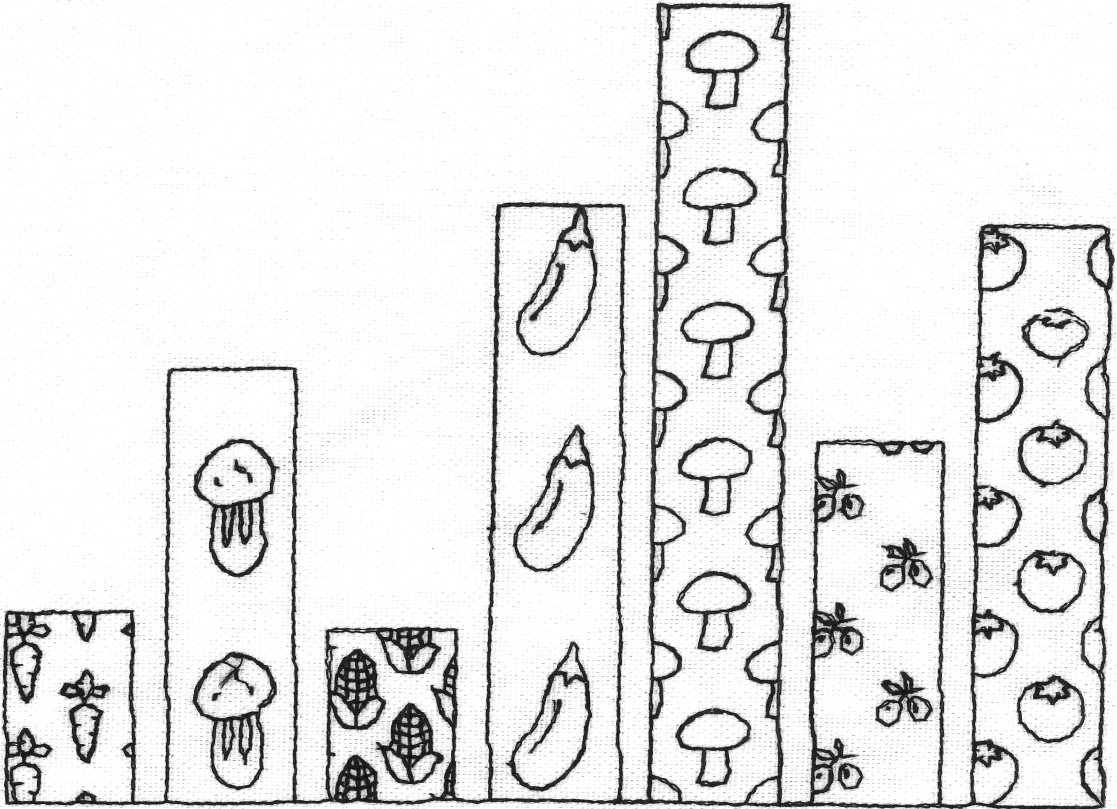}
             \includegraphics[width=0.47\columnwidth]{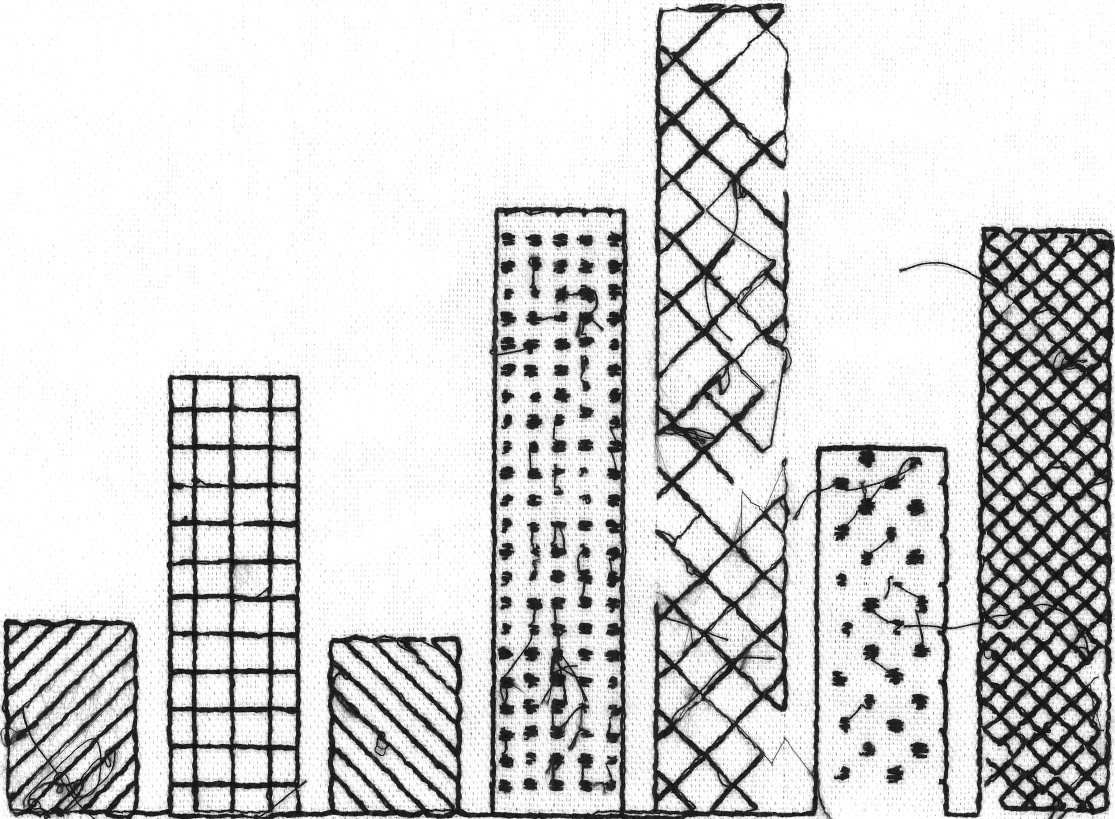}\hfill%
		\includegraphics[width=0.47\columnwidth]{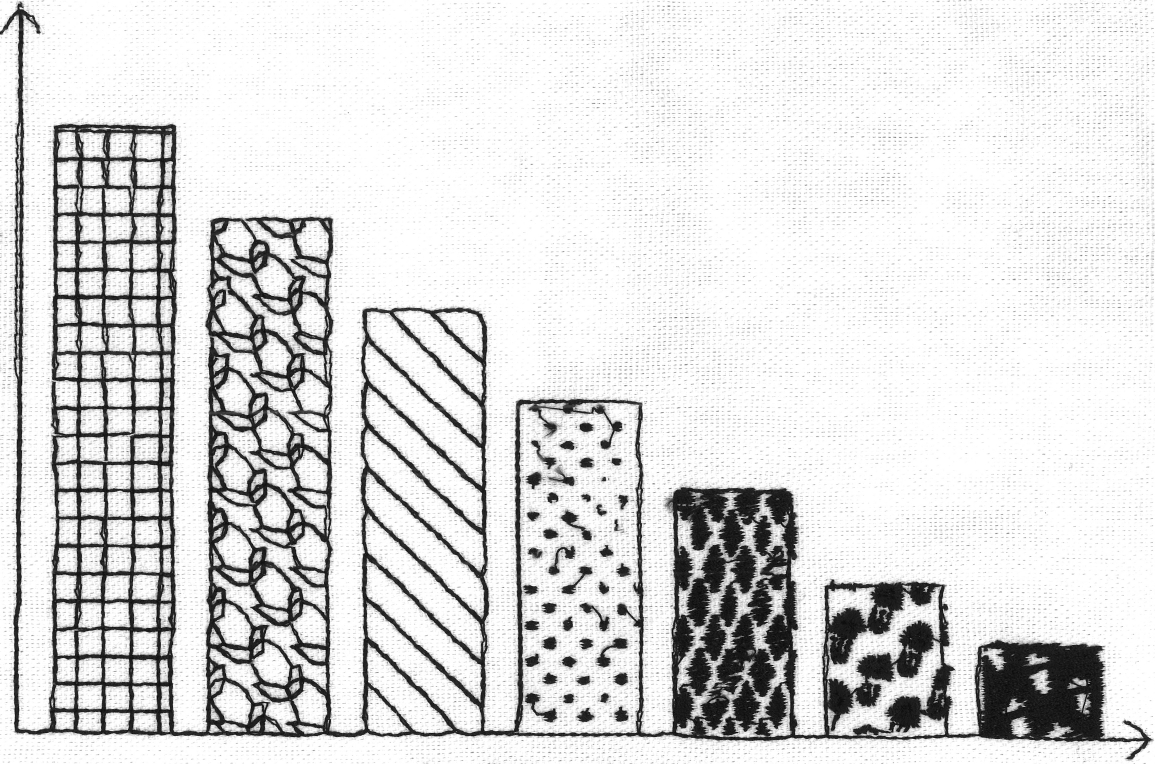}		
    \caption{Some embroidery samples during our experimentation. }
    \label{fig:test_results}
\end{figure}

To identify which textures work better for embroidery we conducted an experimental exploration built upon our previous work \cite{He:2024:DCB,zhong:2020:BWT}. There we had categorized textures into two types: geometric textures (with basic shapes as primitives \inlinevis{-1pt}{1em}{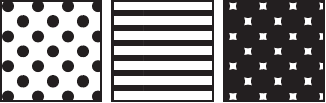}) and iconic textures (with more figurative icons as primitives \inlinevis{-1pt}{1em}{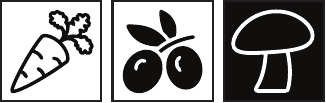}), as well as developed a simple design characterization across different dimensions of textures. We also had collected a set of textured visualization designs from visualization design experts. \autoref{fig:test_results} shows some embroidery samples during our experimentation. While our findings are primarily based on the setup we described in \autoref{sec:machine-embroidery-process}, the fundamental principles are similar across various embroidery machines, thus the conclusions we outline next likely offer broader insights. 

\begin{figure}
    \centering
		(a)~\includegraphics[width=0.9\columnwidth]{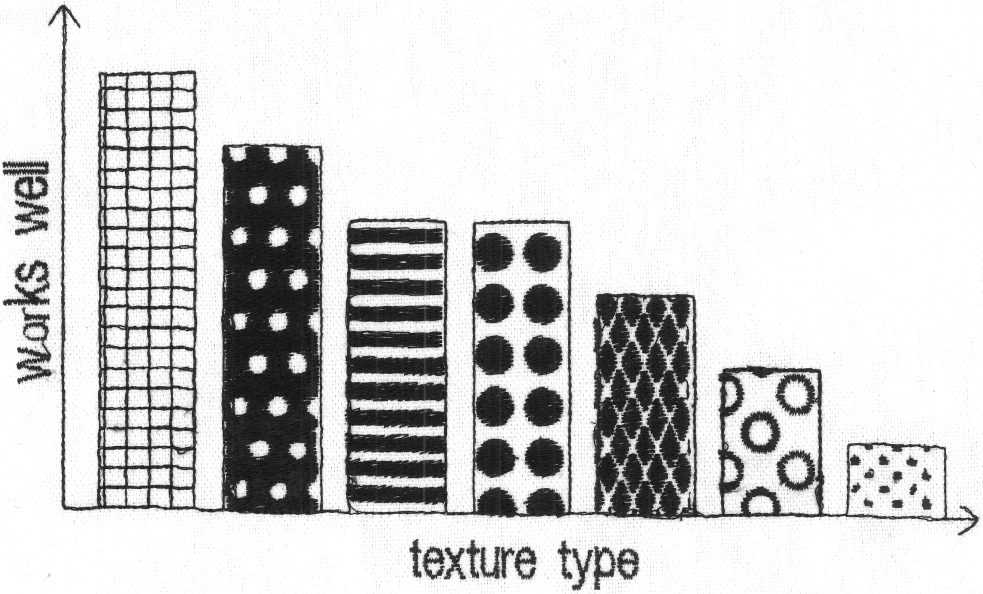}\hfill%
		(b)~\includegraphics[width=0.9\columnwidth]{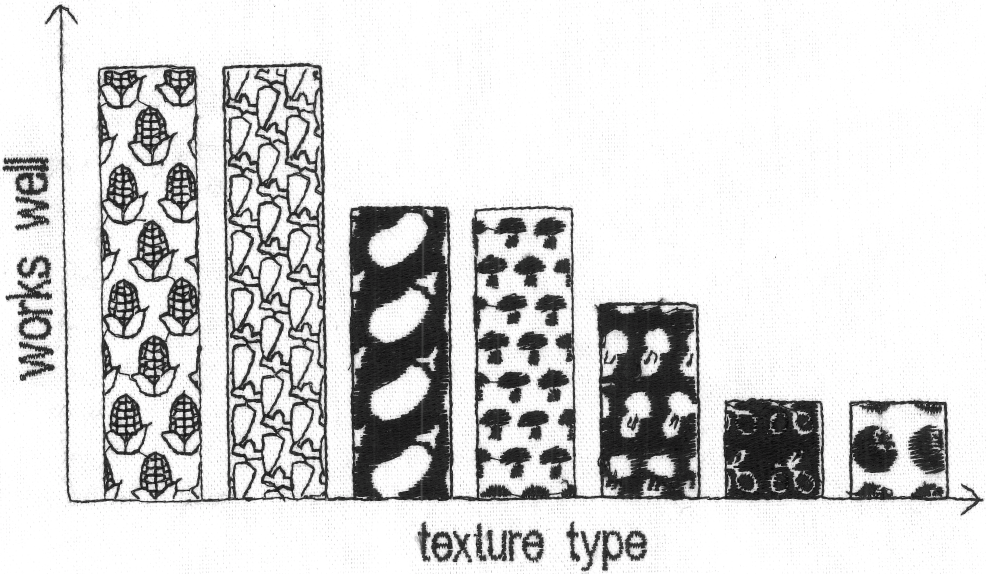}		
    \caption{Two embroidered charts showing the performance of different textures in machine embroidery: (a) for geometric textures, and (b) for iconic textures.}
    \label{fig:texture_performance}
\end{figure}

\begin{figure}
    \centering
		\includegraphics[width=0.9\columnwidth]{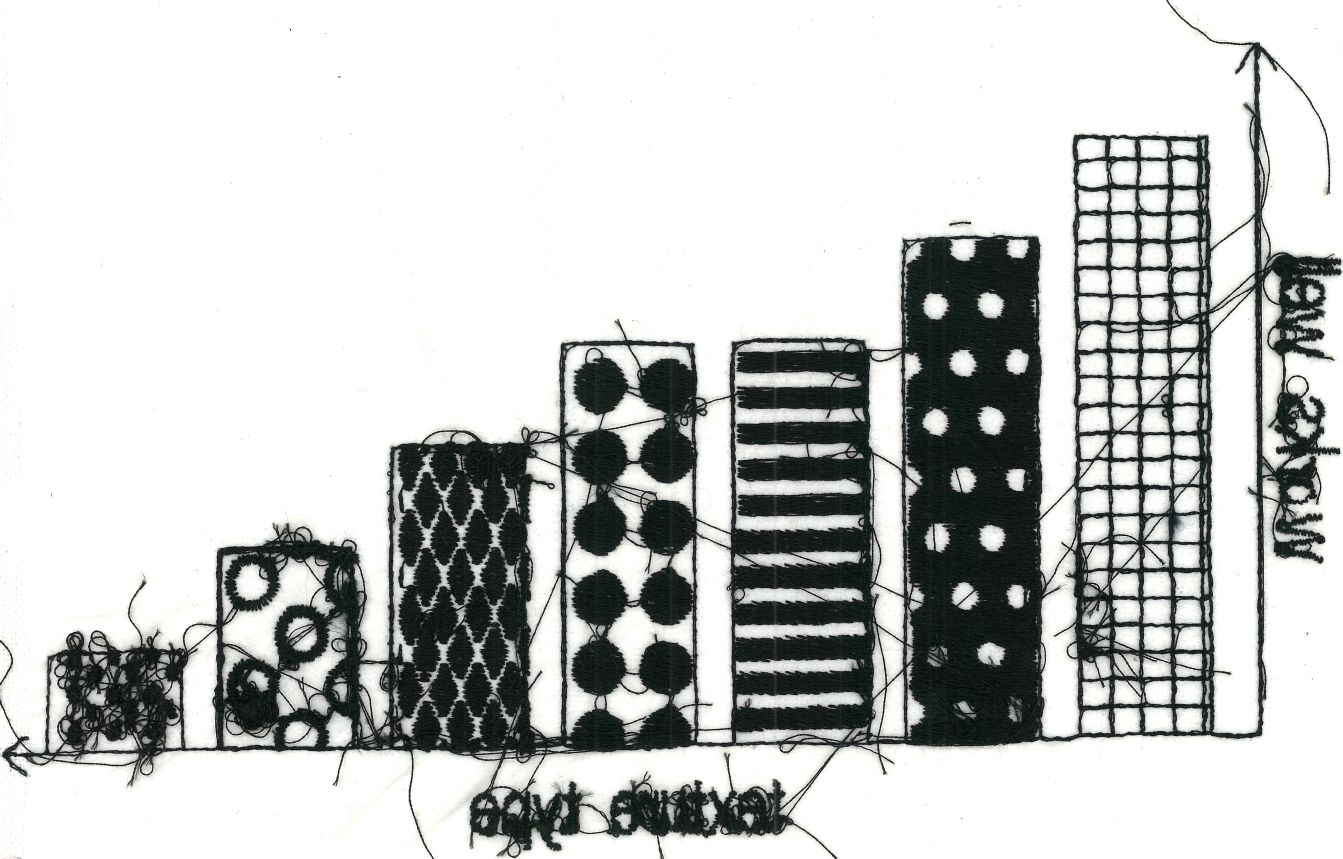}		
    \caption{The reverse side of \autoref{fig:texture_performance}(a). The messy threads indicate that these textures or text elements are problematic to embroider.}
    \label{fig:reverse_texture_performance}
\end{figure}

\textbf{Continuous vs. Scattered Elements.} Embroidery machines perform more efficiently with continuous elements---either as continuous areas or continuous lines (\eg, the left bars in \autoref{fig:texture_performance}(a)). Even detailed line-based icons (\eg, left bars in \autoref{fig:texture_performance}(b)) can be embroidered clearly and accurately. On the other hand, designs with small, scattered areas  (\eg, the right-most bars in \autoref{fig:texture_performance}(a) and (b), respectively) proved problematic for the embroidery process. For example, small dotted textures are prone to cause issues during the embroidery process: their final appearance is not clean (\autoref{fig:texture_performance}(a), right-most bar) and the process often leads to problems as evident in the complex threads on the back side due to the bobbin thread becoming entangled (\autoref{fig:reverse_texture_performance}, left-most bar), while the textures with fewer problems have leaner back sides (visible for the rest of \autoref{fig:reverse_texture_performance}).

\textbf{Simplicity vs. Complexity.} The resolution of the embroidery affects the results of complex patterns. Designs with numerous small details can be difficult to embroider, leading to a loss in clarity and accuracy (\eg, the tomato of the right-most bar in \autoref{fig:texture_performance} lacks the needed detail that separates the main body from the stem, which leads to it being shown essentially as a large dot). Therefore, we observed that simpler patterns lend themselves better to the embroidery process. In our experiments, simple icon designs were translated more effectively into embroidered designs compared to their more complex counterparts.

To illustrate which textures perform better and to facilitate comparisons, we selected some typical examples and visualized their performance using bar charts. The previously mentioned examples in \autoref{fig:texture_performance} compare the performance of the different designs, one for geometric textures and one for iconic textures. In addition, machine embroidery tends to struggle with text, in particular with the small text that is often used for labels or legends in visualizations. In fact, when we look at the reverse side of the embroidery result (\autoref{fig:reverse_texture_performance}), we see that text exhibits similar problems as the textures with scattered elements we mentioned before. Please note that the data shown in \autoref{fig:texture_performance} and~\ref{fig:reverse_texture_performance} are our own subjectively perceived/assessed values of ease of reproduction, they are not precise measurements.

\section{Embroidering Data of a Family Survey}

To evaluate the embroidery workflow and to investigate its use for visualizing personal data, we embroidered the results of a family survey using visualization with black-and-white textures.

\textbf{Data Collection.} The last two authors conducted a survey on their extended family members' preferences for seven vegetables (carrots, celery, corn, eggplant, mushrooms, olives, and tomatos). We asked each family member to rate each of the seven vegetables on a scale of 1 to 5 (1 = ``I don't eat it at all,'' 2 = ``I can eat it if necessary,'' 3 = ``it's ok, I neither like nor dislike it, ``4 = ``I like it,'' 5 = ``it's among my most favorite vegetables''). 11 family members participated in this survey. We calculated the average rating score for each vegetable, and present the result in \autoref{tab:family_survey_scores}.

\begin{table}[b]
\caption{The average scores of the family members' preference towards each of the seven vegetables.}
\label{tab:family_survey_scores}
\centering
\begin{tabular}{c@{\hspace{8pt}}c@{\hspace{8pt}}c@{\hspace{8pt}}c@{\hspace{8pt}}c@{\hspace{8pt}}c@{\hspace{8pt}}c}
\toprule
 carrots                   & celery                   & corn                     & eggplant                 & mushrooms                 & olives                    & tomatos                   \\
\midrule
4.33 & 2.33 & 4.11 & 2.78 & 4.00 & 2.56 & 3.89 \\
\bottomrule
\end{tabular}
\end{table}

\textbf{Visual Representation.} We created a bar chart titled ``How much does my family like vegetables'' to display the results. The $x$-axis represents the type of vegetable, while the $y$-axis indicates the degree of liking according to the mentioned scale. We selected the texture design from a collection of visualization designs with black-and-white textures, gathered from visualization design experts. We chose this particular design not just because it was composed of lines---making it well-suited for the embroidery process---but also because it appealed to a particular family member. In addition, iconic textures can have semantic association, thereby freeing the chart from the need for labels and thus making it extremely suitable for data embroidery. 

\textbf{Embroidery.} To convert the chart into an embroidery, we followed the embroidery process as detailed in \autoref{sec:machine-embroidery-process}. The entire embroidery process took approximately 2 hours, including troubleshooting. \autoref{fig:teaser}(a) shows the data embroidery result. We sewed the embroidered piece onto a canvas bag (see \autoref{fig:teaser}(b)). 
%
%
By integrating family data into such an everyday item that is also used for grocery shopping, this data embroidery can act as a daily reminder of the family's preferences.

\section{Discussion and Future Work}
Given the monochromatic nature of black-and-white textures and, in particular, the semantic association of iconic textures, we envision that they have the potential to become an important visual channel for data embroidery. Our study represents a first step in exploring this technique and we provide our experiences for future embroidery work. Such future work could focus more systematically on evaluating the impact of data embroidery with black-and-white textures in terms of both efficiency and aesthetics \cite{He:2023:BVS}.

\begin{figure}
    \centering
		\includegraphics[width=1\columnwidth]{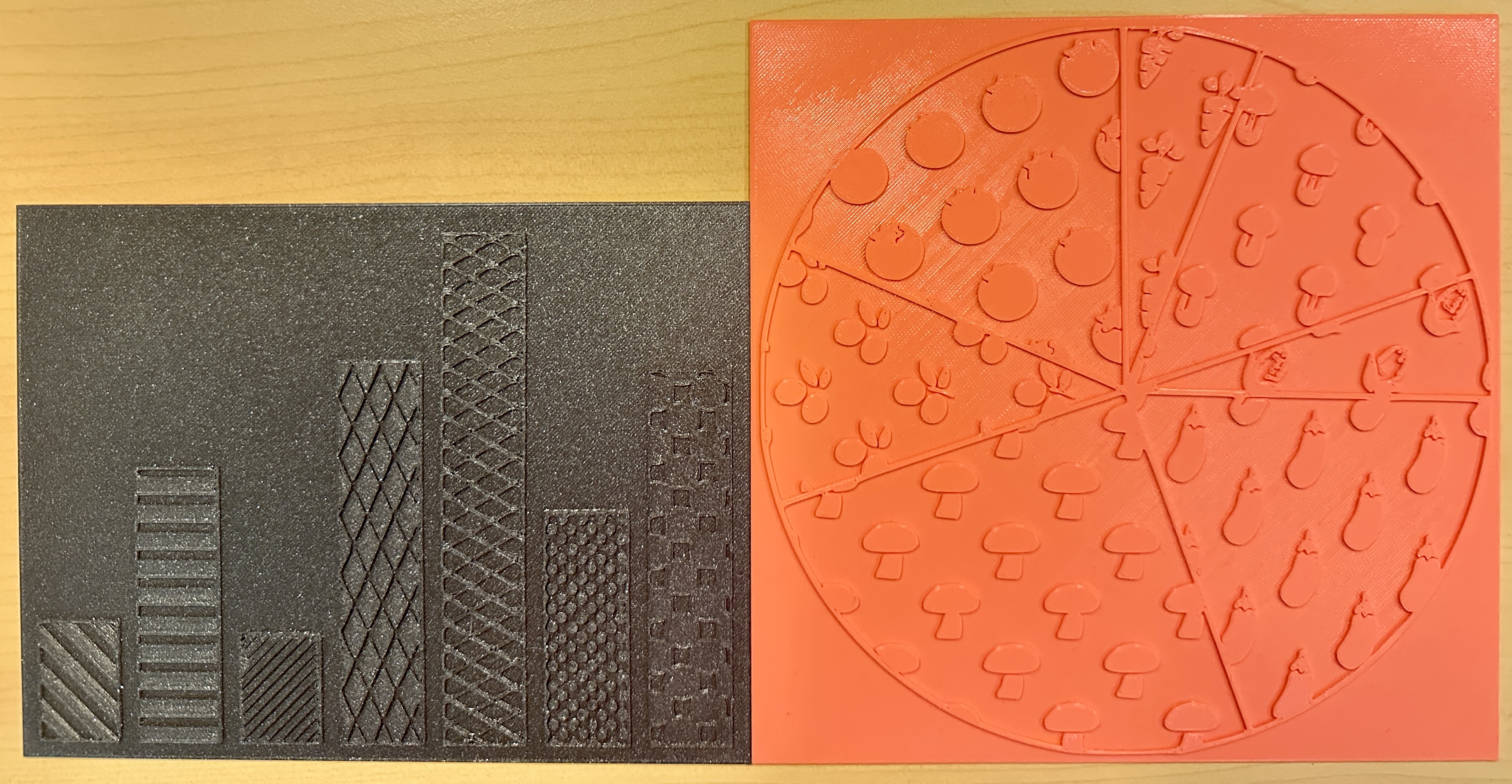}\hfill%
    \caption{Two 3D printed textured charts, one with geometric textures, and another with iconic textures.}
    \label{fig:3D_printed_charts}
\end{figure}

While our work primarily focuses on embroidery, visualizations featuring black-and-white textures can also be easily created physically through a variety of other methods such as 3D printing, laser engraving, vinyl cutting, or embossing. 
We experimented with 3D printing of textured visualizations and produced two charts—one with geometric textures and another with iconic textures (see \autoref{fig:3D_printed_charts}). Compared to embroidery, setting up a 3D printer is simpler and less prone to errors. The process follows basic 3D printing steps: converting the image file (\texttt{PNG} or \texttt{SVG}) to \texttt{STL} using stand-alone or online converters, slicing the STL model to \texttt{GCODE} using a slicer, and, ultimately, importing the \texttt{GCODE} to the 3D printer for printing. The printing process was error-free---textured pieces are not different from any other 3D print.

Our exploration of physicalization using monochrome charts can also open up an interesting additional avenue for future investigation: studying their potential used by visually impaired individuals. Given their tactile nature, these charts could potentially serve as valuable tools for data representation for this population.


\acknowledgments{We thank all visualization experts who provided us with textured visualization designs, as their invaluable insights and expertise served as a crucial cornerstone of our work. We also thank Romain Di Vozzo and Fablab UPSaclay (\href{https://fablabdigiscope.universite-paris-saclay.fr/}{\texttt{fablabdigiscope\discretionary{}{.}{.}universite\discretionary{}{-}{-}paris\discretionary{}{-}{-}saclay\discretionary{}{.}{.}fr}}) for providing us with equipment, training, and technical support.}

\section*{Images/graphs/plots/tables/data license/copyright}
We as authors state that all of our own figures, graphs, plots, and data tables in this article (\ie, those not marked) are and remain under our own personal copyright, with the permission to be used here. We also make them available under the \href{https://creativecommons.org/licenses/by/4.0/}{Creative Commons At\-tri\-bu\-tion 4.0 International (\ccLogo\,\ccAttribution\ \mbox{CC BY 4.0})} license and share them at \href{https://osf.io/f5b6d/}{\texttt{osf.io/f5b6d/}}.

\bibliographystyle{abbrv-doi-hyperref-narrow}
\bibliography{abbreviations,template}


\end{document}